\documentclass[%
aps,
twocolumn,
groupedaddress,
showkeys,
amsmath,
amssymb,
]{revtex4-1}
\usepackage{graphicx}% Include figure files
\usepackage{dcolumn}% Align table columns on decimal point
\usepackage{bm}% bold math
\usepackage{caption}
\usepackage{subfloat}
\usepackage{subfig}
\usepackage{float}
\usepackage{color}
\begin{document}

% Use the \preprint command to place your local institutional report
% number in the upper righthand corner of the title page in preprint mode.
% Multiple \preprint commands are allowed.
% Use the 'preprintnumbers' class option to override journal defaults
% to display numbers if necessary
%\preprint{}

%Title of paper
\title{A simple Tight-Binding Approach to Topological Superconductivity in Monolayer MoS${}_{2}$}

\author{H. Simchi}
\email{simchi@alumni.iust.ac.ir}
\affiliation{Department of Physics, Iran University of Science and Technology, Narmak, Tehran 16844, Iran}
\affiliation{Semicondutor Technology Center, Tehran 19575-199, Iran}

\date{\today}

\begin{abstract}
Monolayer molybdenum disulfide (MoS${}_{2}$) has a honeycomb crystal structure.  Here, with considering the triangular sublattice of molybdenum atoms, a simple tight-binding Hamiltonian is introduced (derived) for studying the phase transition and topological superconductivity in MoS${}_{2}$ under uniaxial strain. It is shown that spin-singlet $p+ip$  wave phase is a topological superconducting phase with nonzero Chern numbers. When the chemical potential is greater (smaller) than the spin-orbit coupling (SOC) strength, the Chern number is equal to four (two) and otherwise it is equal to zero.  Also, the results show that, if the superconductivity energy gap is smaller than the SOC strength and the chemical potential is greater than the SOC strength, the zero energy Majorana states exist. Finally, we show the topological superconducting phase is preserved under uniaxial strain.

\end{abstract}


\maketitle

\section{Introduction}

A monolayer of molybdenum disulfide (MoS${}_{2}$), which has a honeycomb crystal structure, is a prototypical transition-metal dichalcogenide (TMD). Its phonon-limited room temperature mobility is dominated by optical deformation potential and polar optical scattering versus Frohlich interaction\cite{R1}. The dominating deformation potentials originate from couplings to the intervalley LO and homopolar phonons\cite{R1}. It should be noted that due to the valley degeneracy in the conduction band, both intervalley and intravalley scattering of carriers should be considered for calculating the electron-phonon coupling constant\cite{R1}. Roldan et al., have studied the origin of superconductivity in heavily doped MoS${}_{2}$ by considering the electron-phonon and electron-electron interactions\cite{R2}.  They have shown that, the intravalley (intervalley) electron-phonon coupling constant is equal to -0.36 (-0.13)\cite{R2}. It has been shown that heavily gated thin films of MoS${}_{2}$ become superconductive\cite{R3, R4}. The Rashba spin-orbit coupling (RSOC) was arise in the experiment due to the presence of the strong gating field of the order of 10 MeV/cm\cite{R5,R6,R7}. The RSOC induces two superconducting phases which can be topologically nontrivial\cite{R8}. Lu et al., have shown that the Zeeman field, which is originated from intrinsic SOC induced by breaking in-plane inversion symmetry, pins the spin orientation of the electrons to the out-of-plane direction and protects the Ising superconductivity in gated MoS${}_{2}$\cite{R9}. Also, it has been shown that how the spin-triplet p-wave pairing symmetry effects on the superconducting excitations\cite{Rten}. \\

From civil practical point of view, increasing the critical temperature of superconductors ($T_c$) is a necessary condition. For increasing $T_c$, both the density of states (DOS) and the vibration frequency of a superconducting material should be high \cite{R11}. For non-metallic materials or materials which have low DOS, applying strain and doping can be an effective method to induce superconductivity\cite{R12}. The elastic bending modulus of monolayer MoS${}_{2}$ has been studied and it has been shown that the binding modulus is equal to 9.61 eV\cite{R13}. Woo et al., have used a first-principle approach and studied the elastic properties of layered two-dimensional materials\cite{R14}. They have found that the Poisson's ratios of graphene, $h$-BN, and 2H-MoS${}_{2}$ along out-of-plane direction are negative, near zero, and positive, respectively, whereas their in-plane Poisson's ratio are all positive\cite{R14}. Two important and main crystal deformations are mechanical deformation and curvature of crystal lattice. The bond lengths of TMDs are changed under mechanical deformations and in consequence their electronic structures are changed due to the corrections in the electronic Hamiltonian. However, the curvature of the crystal lattice mixes the orbital structure of the electronic Bloch bands. Pearce et al., have presented an effective low energy Hamiltonian for describing the effects of mechanical deformations and curvature on the electronic properties of monolayer TMDs\cite{R15}. In sodium intercalated bilayer MoS${}_{2}$, the electron-phonon interaction strength changes and the superconductivity is significantly enhanced due to the strain effect\cite{R16}. Similar work has been reported about calcium doped MoS${}_{2}$ bilayer, recently\cite{R17}. Finally, Kang et al., have reported the discovery of Holstein polarons (a small composite particle of an electron that carriers a cloud of phonons) in a surface-doped layered semiconductor, MoS${}_{2}$\cite{R18} and  He et al., have shown that in a wide range of experimentally accessible regime where the in-plane magnetic field is higher than the Pauli limit field but lower than second critical magnetic field, a 2H-structure monolayer NbSe${}_{2}$ or similarly TaS${}_{2}$ becomes a nodal topological superconductor\cite{R19}.
\\

In this paper, we study the spin-singlet $p+ip$  topological superconductivity in monolayer MoS${}_{2}$ using a simple tight-binding Hamiltonian when the crystal is (is not) under uniaxial strain.

Xiao et al., have considered the molecular orbitals $d_{z^2}$ and $1/\sqrt{2}(d_{x^2-y^2}+i\tau d_{xy})$ as basis functions of conduction and valence bands, respectively ($\tau=\pm1$ is the valley index).They have shown that, to first order in $k$, the $C_{3h}$ symmetry dictates a two-band  $\overrightarrow{k}\cdot\overrightarrow{p}$ Hamiltonian\cite{R33}. Cappelluti et al., have shown that, the $4d_{3z^2-r^2} (\%82)$, $3p_{x}$,$3p_{y} (\%12)$ and other orbitals$(\%6)$ contribute in  constructing the minimum of the conduction band (MCB)\cite{R32}.  Roldan et al., have studied the superconductivity effect in heavily doped MoS${}_{2}$ by considering the $d_{3z^2-r^2}$  orbital as main orbital component\cite{R2}. The possible topological superconductivity phase\cite{R8} and strongly enhanced superconductivity\cite{R12} in MoS${}_{2}$ have been studied. In the both studies, the $d_{3z^2-r^2}$ orbital is considered as the main orbital component\cite{R8,R12}.Therefore, we consider a single band Hamiltonian model including intrinsic SOC (ISOC) and RSOC near $K$ and $K^\prime$ points and show that there are two band inversion (nodal) points which separate unoccupied states from occupied ones in phase diagram. Using the threefold rotation symmetry of the crystal and the properties of rotation fixed points $\mathrm{\Gamma }$ $,\ K$ and $K^\prime$, we calculate the Chern number\cite{R20}. It is shown that when the chemical energy is greater (smaller) than the ISOC strength, the Chern number is equal to four (two) and otherwise it is equal to zero.  Since, the $4d_{z^2}$ orbital is the dominant component of the states we only consider the Mo-layer with triangular lattice structure for doing the numerical calculations. The Mo-layer has both flat and zigzag edges but we consider a zigzag nanoribbon of Mo-layer and introduce a simple tight-binding Hamiltonian for studying the topological superconductivity in MoS${}_{2}$. It is shown that when the superconductivity energy gap is smaller than ISOC and the chemical potential is greater than ISOC, the zero energy Majorana states exist. Finally, we show that the topological superconducting phase is preserved under uniaxial strain. Using a simple tight binding method and crystal symmetry for studying the topological properties and justifying the effect of uniaxial strain on the topological properties are the novelties of the article in comparison with the other published articles\cite{R8}. The structure of article is as follows: The analytical calculations are presented in section II. In section III, the tight-binding Hamiltonian is introduced. The results are explained and discussed in section IV. In section V, the summary is presented.

\section{Analytical calculations}\label{sec:analyticalcalculations}

\subsection{Without applying strain}
Many tight-binding Hamiltonian models have been proposed for monolayer MoS${}_{2}$ \cite{R21,R22,R23,R24,R25,R26,R27}. Since, the $4d_{z^2}$ orbital is the dominant component near $K$ and $K^\prime$ points, a single band general Hamiltonian in the basis of $(c_{\boldsymbol{k}\uparrow },c_{\boldsymbol{k}\downarrow })$ can be written as\cite{R8,R27}
\begin{equation} \label{GrindEQ__1_} 
H_0\left(\boldsymbol{k}+\epsilon \boldsymbol{K}\right)=(\frac{{\left|\boldsymbol{k}\right|}^2}{2m}-\mu){\sigma }_0 +{\alpha }_R\boldsymbol{g}\left(\boldsymbol{k}\right){\color{red}\cdot} \boldsymbol{\sigma }+\epsilon {\beta }_{so}{\sigma }_z 
\end{equation} 
where, $c_s$ is electron annihilation operator, $s=\uparrow /\downarrow $ denotes spin, $\epsilon =\pm 1$ is the valley index, $m$ is the effective mass of the electrons and $\mu $ is the chemical potential measured from the conduction band minimum when SOC is omitted. $\boldsymbol{g}\left(\boldsymbol{k}\right)=(k_y,-k_x,0)$ and ${\alpha }_R$ are Rashba vector and RSOC strength coefficient, respectively. Finally, ${\beta }_{so}$ is ISOC strength coefficient and $\boldsymbol{\sigma }$\textbf{ }denotes the Pauli matrices. ${\sigma }_0$ is unit matix.\\

The nearest-neighbor spin-singlet superconducting pairing amplitudes are proportional to           $e^{-{\beta }_{ij}}\left\langle c_{i,\downarrow }c_{j,\uparrow }-c_{i,\uparrow }c_{j,\downarrow }\right\rangle $ in the two dimensional representation\cite{R8}. Here, ${\beta }_{ij}$is the nearest-neighbor pairing of the spin-singlet $p$-wavelike phase (see Fig.1c)\cite{R8}. One can substitute the Fourier transformation of $c_{i,s}$ in the relation and shows
\begin{equation} \label{GrindEQ__2_} 
\sum_{<ij>}{\left(c_{i,\downarrow }c_{j,\uparrow }-c_{i,\uparrow }c_{j,\downarrow }\right)=\sum_{k}{2{\mathrm{cos} \left(\boldsymbol{k}{\color{red}\cdot} {\boldsymbol{\tau }}_{\boldsymbol{i}}\right)\ }c_{-k,\downarrow }c_{k,\uparrow }}} 
\end{equation} 
where, ${\boldsymbol{\tau }}_{\boldsymbol{i}}$\textbf{ }are the bonding vectors of Mo-atoms i.e., $\tau_{1}=a(1,0)$\textbf{, }${\tau_{2}}\boldsymbol{=}a\boldsymbol{(-}\frac{\boldsymbol{1}}{\boldsymbol{2}},\frac{\sqrt{\boldsymbol{3}}}{\boldsymbol{2}}\boldsymbol{)}$\textbf{, }${\tau_{3}}\boldsymbol{=}a\left(\frac{\boldsymbol{-}\boldsymbol{1}}{\boldsymbol{2}},\frac{\boldsymbol{-}\sqrt{\boldsymbol{3}}}{\boldsymbol{2}}\right)$\textbf{ }with lattice constant a=3.16 Angstrom($\AA$)(Fig.1b) and $k=k_{x}+ik_{y}$ \cite{R8}. Therefore\cite{R8},
\begin{widetext}
\begin{equation}
\Delta \left(k\right)=2\{{\mathrm{cos} \left({ak}_x\right)\ }+e^{\frac{i2\pi }{3}}\left[{\mathrm{cos} \left(-\frac{ak_x}{2}\right)\ }+{\mathrm{cos} \left(\frac{a\sqrt{3}}{2}k_y\right)\ }\right]+e^{\frac{i4\pi }{3}}\left[{\mathrm{cos} \left(-\frac{ak_x}{2}\right)\ }+{\mathrm{cos} \left(\frac{-a\sqrt{3}}{2}k_y\right)\ }\right]\}
\end{equation} 
\end{widetext}
                                                                                                                                             
By using Eqs. (1) and (3), in the basis of $(c_{k,\uparrow },c_{k,\downarrow },c^\dagger_{-k,\uparrow },c^\dagger_{-k,\downarrow })$, the general Bogoliubov-de Genes (BdG) Hamiltonian can be written as\cite{R8}
\begin{widetext}
\begin{equation} \label{GrindEQ__4_} 
H_{BdG}=\left( \begin{array}{cc}
 \begin{array}{cc}
{\varepsilon }_k+\epsilon \mathrm{\ }{\beta }_{so} & {i\alpha }_R(k_x{-ik}_y) \\ 
{-i\alpha }_R(k_x{+ik}_y) & {\varepsilon }_k-\epsilon \mathrm{\ }{\beta }_{so} \end{array}
 &  \begin{array}{cc}
0 & \Delta (k) \\ 
-\Delta (k) & 0 \end{array}
 \\ 
 \begin{array}{cc}
0 & -{\Delta }^*(k) \\ 
{\Delta }^*(k) & 0 \end{array}
 &  \begin{array}{cc}
{-(\varepsilon }_k+\epsilon \mathrm{\ }{\beta }_{so}) & -{i\alpha }_R(k_x{+ik}_y) \\ 
{i\alpha }_R(k_x{-ik}_y) & -({\varepsilon }_k-\epsilon \mathrm{\ }{\beta }_{so}) \end{array}
 \end{array}
\right) 
\end{equation} 
\end{widetext}
where, ${\varepsilon }_k=\frac{k^2}{2m}-\mu $. It should be noted that $(c_{k,\uparrow },c_{k,\downarrow })$ and ($c^\dagger_{-k,\uparrow },c^\dagger_{-k,\downarrow })$ are basis vectors of electrons and holes, respectively. One can easily show that the eigenvalues of $H_{BdG}$ are
\begin{equation} \label{GrindEQ__5_} 
E^2=({\varepsilon }^2_k+{\beta }^2_{s0}+{\alpha }^2_Rk^2+{\Delta }^2)\pm 2{\left[\left({\beta }^2_{s0}+{\alpha }^2_Rk^2\right){\varepsilon }^2_k+{\Delta }^2{\beta }^2_{s0}\right]}^{1/2} 
\end{equation} 
Using the Tailor expansion of Eq.(3) near $K(K^\prime)$- point, it can be shown that
\begin{equation} \label{GrindEQ__6_} 
\Delta \left(\boldsymbol{k}+\epsilon \boldsymbol{K}\right)\approx \gamma \epsilon (k_x+ik_y) 
\end{equation} 
where, $\gamma =4\sqrt{3}$\cite{R8}. However, the points $\mathrm{\Gamma }$ $,\ K$ and $K^\prime$ are rotation fixed points of threefold rotation\cite{R25}. Therefore, we can study the change of the Chern number by the change of the rotation eigenvalues at $ K$ and $K^\prime$ points ($k_x \&\ k_y\to 0)$\cite{R28}. However, the eigenvalues of Eq.\eqref{GrindEQ__4_} at $K$-point are

\noindent $E_1=-{\beta }_{so}+\mu $, $E_2=-{\beta }_{so}-\mu $, $E_3=-{\beta }_{so}+\mu $, and $E_4={\beta }_{so}+\mu $ with eigenfunctions       ${\psi }_1={\left(1,0,0,0\right)}^T$, ${\psi }_2={\left(0,1,0,0\right)}^T$, ${\psi }_3={\left(0,0,1,0\right)}^T$, and ${\psi }_4={\left(0,0,0,1\right)}^T$, respectively.

\noindent Similarly, for $K^\prime$-point we have

\noindent $E^\prime_1=-{\beta }_{so}-\mu $, $E^\prime_2=-{\beta }_{so}+\mu $, $E^\prime_3={\beta }_{so}+\mu $, and $E^\prime_4={-\beta }_{so}+\mu $ with eigenfunctions       ${\psi }^\prime_1={\left(1,0,0,0\right)}^T$, ${\psi }^\prime_2={\left(0,1,0,0\right)}^T$, ${\psi }^\prime_3={\left(0,0,1,0\right)}^T$, and ${\psi }^\prime_4={\left(0,0,0,1\right)}^T$, respectively.

\noindent It should be noted that ${\psi }_1$ and ${\psi }^\prime_1$  (${\psi }_2\ and \ {\psi }^\prime_2$) are eigenfunctions of spin-up (down) electrons, but ${\psi }_3$  and ${\psi }^\prime_3$ (${\psi }_4 and \ {\psi }^\prime_4$) are eigenfunctions of spin-up (down) holes in the basis of $(c_{k,\uparrow },c_{k,\downarrow },c^\dagger_{-k,\uparrow },c^\dagger_{-k,\downarrow })$.

\subsection{With applying strain}
 It has been shown that the low-energy d-bands effective Hamiltonian around$\ K$-point and in the space of $(d^{\epsilon }_{z^2,\boldsymbol{k},s},d^{\epsilon }_{\epsilon 2,\boldsymbol{k},s})$ can be written as \cite{R15}
\begin{equation} \label{GrindEQ__7_} 
H^{E,\epsilon }_{k,eff}=\left( \begin{array}{cc}
{\Delta }_c+\beta {\left|\boldsymbol{k}\right|}^2 & v\boldsymbol{k}+\kappa {\boldsymbol{k}}^{+2} \\ 
v{\boldsymbol{k}}^++\kappa {\boldsymbol{k}}^2 & {\mathrm{\Delta }}_v+\alpha {\left|\boldsymbol{k}\right|}^2+2\epsilon {\beta }_{so}s_z \end{array}
\right) 
\end{equation} 
where, for MoS${}_{2}$, ${\Delta }_c=1.78$ eV (conduction band edge), ${\mathrm{\Delta }}_v=-0.19$ eV (valence band edge), $v=2.44$ eV(group velocity), $\beta =0.21eV {\AA}^2$(effective mass), $\alpha =0.71eV{\AA}^2$(effective mass), $\kappa =0.32eV{\AA}^2$(higher order in momentum trigonal corrections), ${\beta }_{so}$=0.06 eV, and $s_z=\pm 1$\cite{R15}. Of course, the cubic correction terms are omitted, here. \\

The strain tensor of a two dimensional membrane is given by\cite{R13, R14, R15}
\begin{equation} \label{GrindEQ__8_} 
{\epsilon }_{ij}=\frac{1}{2}\{{\partial }_iu_j+{\partial }_ju_i+\left({\partial }_ih\right)\left({\partial }_jh\right)\} 
\end{equation} 
where, \textbf{u}$(\boldsymbol{r})$ and $h(\boldsymbol{r})$ are in-plane and out-of-plane deformation vectors, respectively such that a generic atom at position $\boldsymbol{r}$ is shifted to ${\boldsymbol{r}}^{\boldsymbol{'}}=\boldsymbol{r}+\boldsymbol{u}\left(\boldsymbol{r}\right)+\widehat{\boldsymbol{z}}h(\boldsymbol{r})$ under the deformation. Without curvature, the Eq.\eqref{GrindEQ__7_} can be written as\cite{R15}

\begin{widetext}
 \begin{equation}
H^{E,\epsilon }_{k,eff}=\left( \begin{array}{cc}
{\Delta }_c+\beta {\left|\boldsymbol{k}\boldsymbol{+}{\eta }_2F^{\epsilon }\right|}^2+{\delta }_1D+{\delta }_3D^2+s_zD(\delta {\beta }_{so1}) & v\boldsymbol{k}\boldsymbol{+}{\eta }_1F^{\epsilon }+\kappa {\left({\boldsymbol{k}}^{\boldsymbol{+}}+{\eta }_4F^{\epsilon +}\right)}^2 \\ 
v{\boldsymbol{k}}^++{\eta }_1F^{\epsilon +}+\kappa {\left(\boldsymbol{k}+{\eta }_4F^{\epsilon }\right)}^2 & {\mathrm{\Delta }}_v+\alpha {\left|\boldsymbol{k}\boldsymbol{+}{\eta }_3F^{\epsilon }\right|}^2++{\delta }_2D+{\delta }_4D^2+s_zD(\delta {\beta }_{so2}) \end{array}
\right) 
\end{equation} 
\end{widetext}

where, $D=Tr[{\epsilon }_{ij}]$ and  $F^{\epsilon }=({\epsilon }_{yy}-{\epsilon }_{xx}+i\epsilon 2{\epsilon }_{xy})$ \cite{R15}. The values of all constants in Eq.(9) are provided in Ref.15 and we do not repeat them, here. From Eq. (9), it can be concluded that the terms $\{\beta {\left|{\eta }_2F^{\epsilon }\right|}^2+{\delta }_1D+{\delta }_3D^2+s_zD\left(\delta {\beta }_{so1}\right)\}$ and $\{$$\alpha {\left|{\eta }_3F^{\epsilon }\right|}^2+{\delta }_2D+{\delta }_4D^2+s_zD\left(\delta {\beta }_{so2}\right)\}$ should be added to Eq.\eqref{GrindEQ__4_} when we want to study the effect of strain on the topological properties at$\ K$ and $K^\prime$ points ($k_x\&\ k_y\to 0).$ We will show that the off-diagonal terms are very small and in consequence they are negligible. It should be noted that it is assumed the crystal is flat and there is no any curvature under applying the deformations.

\section{Tight-binding Hamiltonian}\label{sec:numerical}
As it has been shown, the $4d_{z^2}$ orbital is the dominant component of the states near the CBM and the VBM located at $K$ and $K^\prime$ points \cite{R2, R8, R12}. Therefore, we consider a zigzag nanoribbon of Mo-layer with triangular crystal structure (Fig.1(a)) and use a simple tight-binding Hamiltonian for studying the topological superconductivity in MoS${}_{2}$. Rostami et al., have been shown that if the basis ${\psi }_{Mo}=\left(d_{z^2},d_{x^2-y^2},d_{xy}\right)\ $is used, the Hamiltonian of Mo-Mo hopping can be written in $k$-space as\cite{R29}

\begin{equation} \label{GrindEQ__10_} 
H_{MM}=\left( \begin{array}{ccc}
{\varepsilon }_0 & 0 & 0 \\ 
0 & {\varepsilon }_2 & -i{\beta }_{so,Mo}s_z \\ 
0 & i{\beta }_{so,Mo}s_z & {\varepsilon }_2 \end{array}
\right) +2\sum^3_{i=1}{t^{MM}_i\mathrm{cos}\mathrm{}(\boldsymbol{k}.{\boldsymbol{\tau }}_{\boldsymbol{i}})} 
\end{equation} 

\begin{figure*}[!ht]
\captionsetup{singlelinecheck = false, justification=raggedright}
\includegraphics[width=\columnwidth]{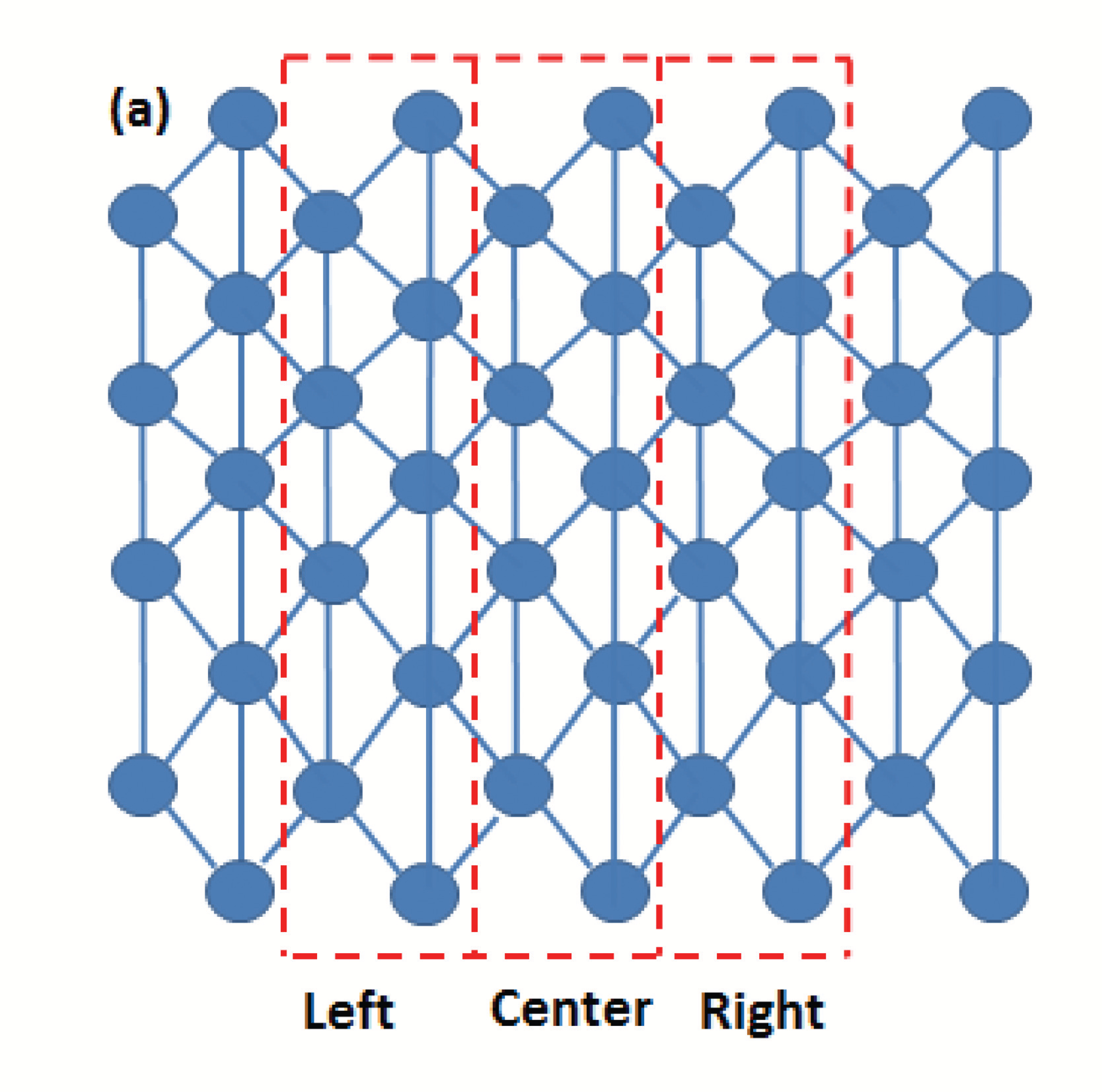}
 \includegraphics[width=\columnwidth]{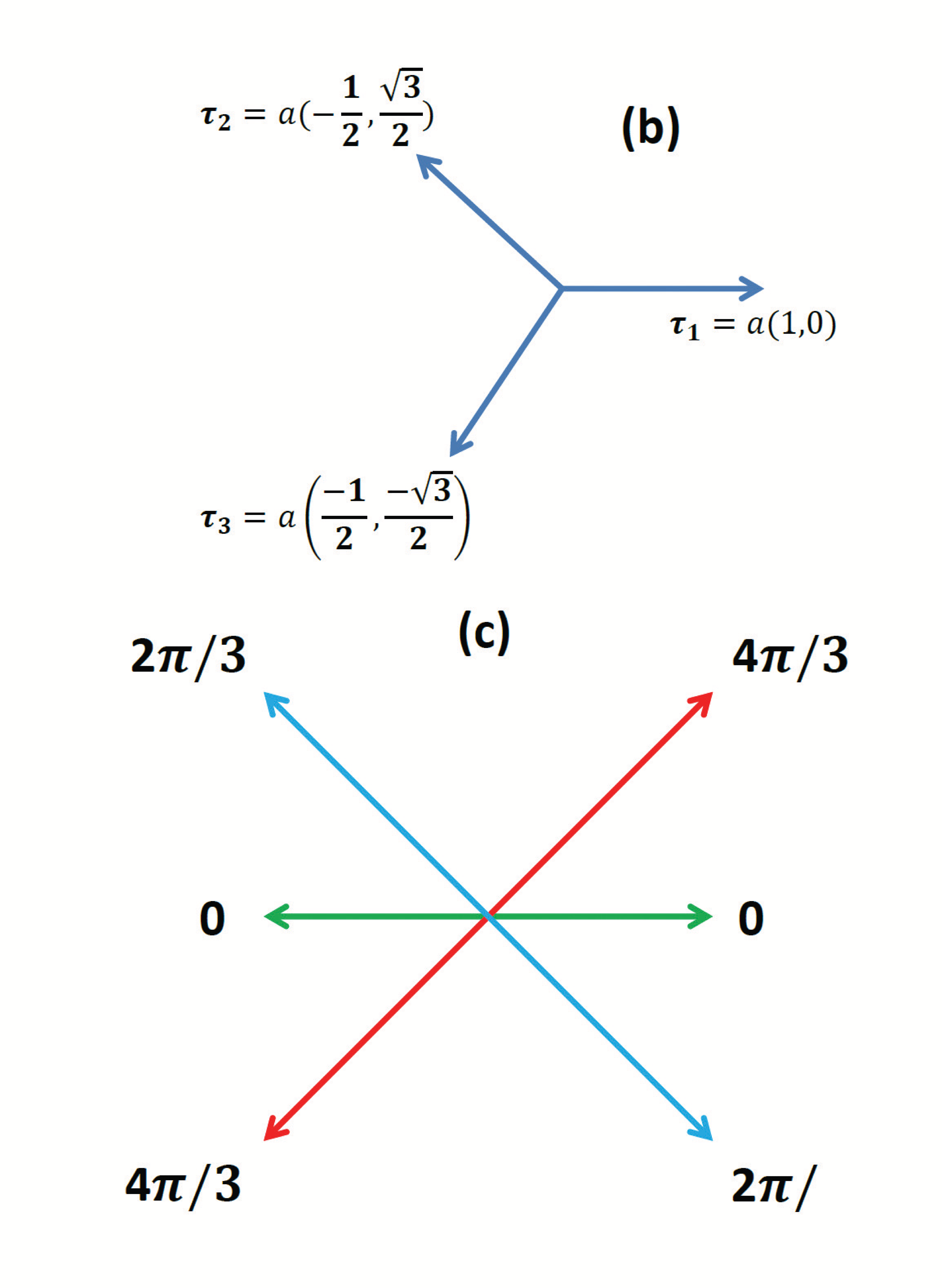}
 \caption{(Color online) (a) Triangular sublattice of monolayer $Mo$-atoms, (b) the bonding vectors of $Mo$-atoms, and (c) the nearest-neighbor pairing phases${\beta}_{ij}$ of the spin-singlet, $p$-wavelike phase\cite{R8}.}
\end{figure*}

where, ${\varepsilon }_0$ and ${\varepsilon }_2$ are on-site energies, and $t^{MM}_i$is hopping matrix in ${\boldsymbol{\tau }}_{\boldsymbol{i}}$\textbf{-}direction(Fig.1(b)).\textbf{ }The all constant values of Eq.\eqref{GrindEQ__10_} are provided in Ref.29 and we do not repeat them, here. In our real-space model, ${\varepsilon }_0=1.282$ eV,  ${\varepsilon }_2=0.864$ eV, and hopping terms $t_i=2t^{MM}_i$ are assumed. It should be noted that the Mo-S and S-S hopping terms are not considered, and in consequence, the values are assumed such that the next results not only provide the correct physics but also confirm the results of analytical calculations.  Fig.1 shows a zigzag nanoribbon of Mo-atoms in triangular lattice structure. Using Eq.\eqref{GrindEQ__10_} in real-space\cite{R29}, one can consider the central, left and right supercells and write their Hamiltonian matrices i.e., $H_{00}$, $H_{01}$, and $H_{01}$, respectively. The energy dispersion curve of the nanoribbon can be found by finding the eigenvalues of the below Hamiltonian
\begin{equation} \label{GrindEQ__11_} 
H^{\uparrow (\downarrow )}=H^{\uparrow (\downarrow )}_{00}+e^{i\boldsymbol{k}.\boldsymbol{A}}H^{\uparrow (\downarrow )}_{01}+e^{-i\boldsymbol{k}.\boldsymbol{A}}H^{\uparrow (\downarrow )}_{10} 
\end{equation} 
where, $\boldsymbol{A}$ is lattice vector of the nanoribbon. Since, $s_z=+1\ (-1)$ for spin-up ($\uparrow $) (down ($\downarrow $)) electrons, we can write the Hamiltonians for both types of spin. In each supercell, the nearest-neighbor pairing term is equal to ${\Delta }_0e^{i{\beta }_{ij}}$ in real-space which the value of  ${\beta }_{ij}$ depends on the direction (${\boldsymbol{\tau }}_{\boldsymbol{i}}$)  as shown in Fig.1(c)\cite{R8}. It is assumed that the paring happens only between the $4d_{z^2}$ orbitals\cite{R8}. Therefore, the paring matrix between two Mo-nearest neighbor atoms can be written as 
\begin{equation} \label{GrindEQ__12_} 
\Delta ={\Delta }_0e^{i{\beta }_{ij}}\left( \begin{array}{ccc}
1 & 0 & 0 \\ 
0 & 0 & 0 \\ 
0 & 0 & 0 \end{array}
\right) 
\end{equation} 
However, the Hamiltonian of holes is equal to $-H^{*e}(-\boldsymbol{k})$ where $H^e(\boldsymbol{k})$ is the Hamiltonian of electrons\cite{R8}. Thus, one can write the BdG-Hamiltonian as follows

\noindent 
\begin{equation} \label{GrindEQ__13_} 
H_{BdG}=\left( \begin{array}{cc}
 \begin{array}{cc}
H^{e,\uparrow } & R \\ 
R^+ & H^{e,\downarrow } \end{array}
 &  \begin{array}{cc}
0 & \Delta  \\ 
-\Delta  & 0 \end{array}
 \\ 
 \begin{array}{cc}
0 & -{\Delta }^* \\ 
{\Delta }^* & 0 \end{array}
 &  \begin{array}{cc}
H^{h,\uparrow } & R^+ \\ 
R & H^{h,\downarrow } \end{array}
 \end{array}
\right) 
\end{equation} 
where, $R$ stands for RSOC which is assumed to be equal to zero in next calculations because we study the superconductivity effect at $K$ ($K\prime$)-point. It should be noted that for adding RSOC effect to the model, one should find its analytical formula in the basis ${\psi }_{Mo}=\left(d_{z^2},d_{x^2-y^2},d_{xy}\right)$.\\

Another simple model can be used to study the effect of chemical potential, respect to the ISOC strength, on the topological properties of MoS${}_{2}$. In this model, only the $d_{z^2}$ orbital is considered, and the term $-\mu +\epsilon {\beta }_{so}$ is placed on the diagonal of $H^{\uparrow (\downarrow )}_{00}$.  For hopping between two Mo-nearest neighbor atoms, the hopping energy $t=(3V_{dd\delta }+V_{dd\sigma })/2$ is used in all directions\cite{R29} and the above explained method is followed again. It should be noted that in calculations $\Delta ={\Delta }_0e^{i{\beta }_{ij}}$ \cite{R8}, $V_{dd\delta }=0.228$ eV, and $V_{dd\sigma }=-0.895$ eV\cite{R29}. 

\section{Results and discussion}\label{sec:results}
The eigenvalues of Eq.\eqref{GrindEQ__4_} at $K$ and $K^\prime$ points were found in section II.  Fig. 2 shows them, schematically. 
\begin{figure}
\captionsetup{singlelinecheck = false, justification=raggedright}
{
    \includegraphics[width=\columnwidth]{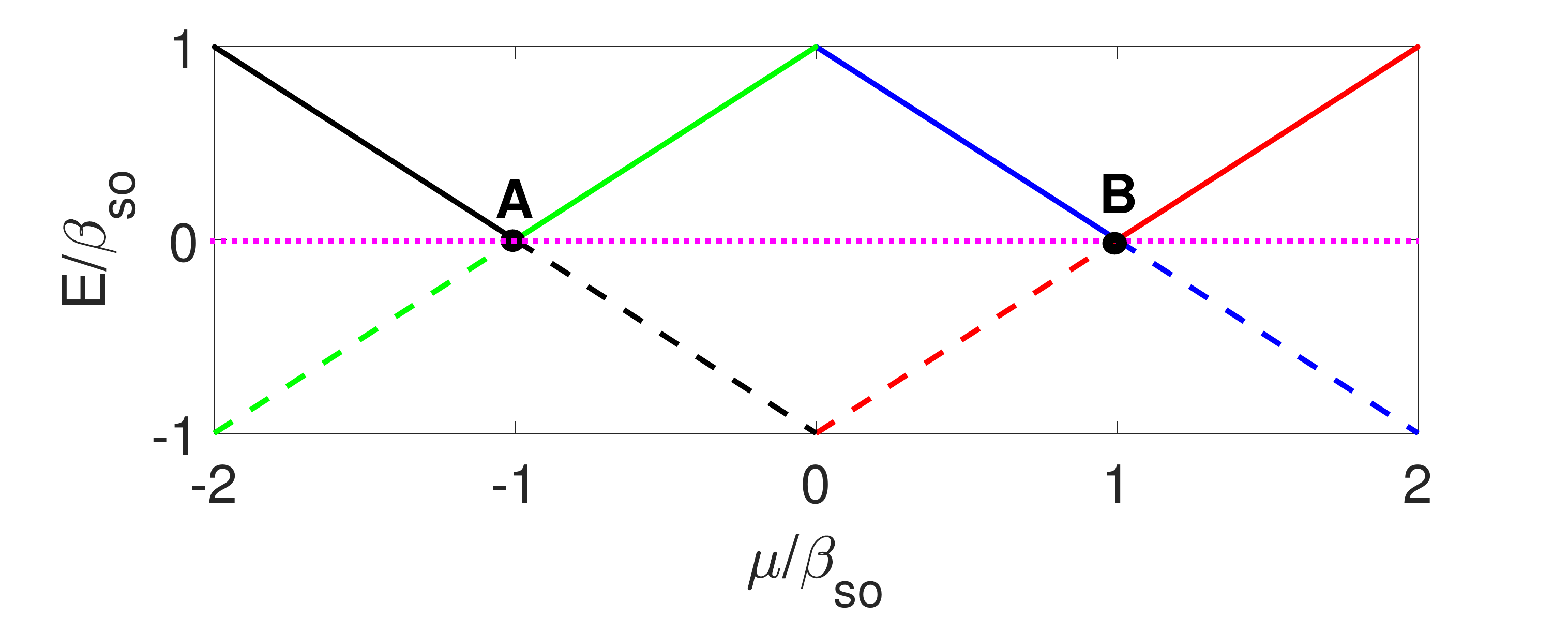}
}
\caption{(Color online) Energy dispersion curve ( called phase diagram) of superconducting MoS${}_{2}$ at $K$($K^\prime$) -point. The dashed and filled lines show the negative and positive energies, respectively. The point A and B stand for band-inversion (nodal) points.The general equation $E/{\beta}_{so}=\pm(1\pm\mu/{\beta}_{so})$ is used, here.}
\end{figure}
\begin{figure}
\captionsetup{singlelinecheck = false, justification=raggedright}
{
    \includegraphics[width=\columnwidth]{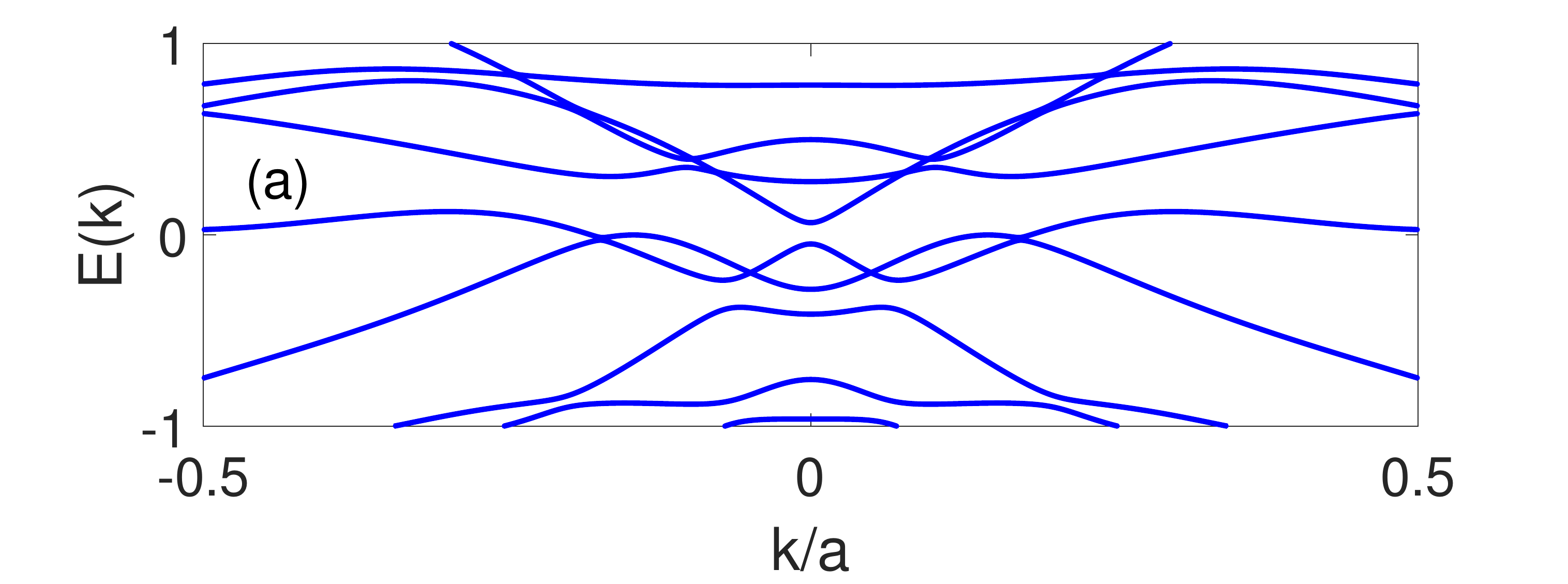}
}
\\

{
    \includegraphics[width=\columnwidth]{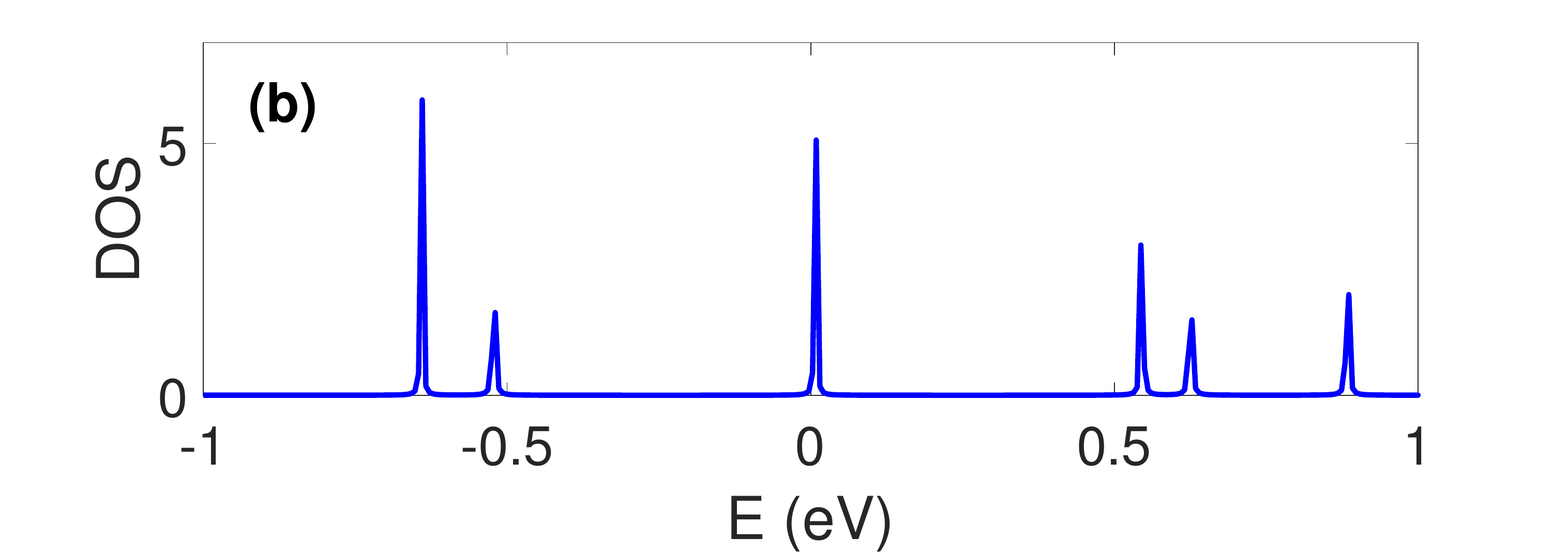}
}
\caption{(a) Energy dispersion curve of a zigzag nanoribon of $Mo$ atoms and (b) density of states ($DOS$) versus energy. It should be noted that $RSOC$ is not considered. Since ${\beta}_{so}=60$ meV, the difference between spin up and down electrons is negligible. Each supercell includes nine Mo atoms.}
\end{figure}
As Fig. 2 shows, the curves of negative energies touch the curves of positive energies at two point A and B which are called band-inversion (nodal) points.However, the points are placed on the line, $E/{\beta }_{so}=0$. Therefore, band closing happens at $K$ and $K^\prime$ points. Since the $K$ and $K^\prime$ points are rotation fixed points, it is possible to label the states at the fixed points by their rotation eigenvalues which are $e^{i\pi \left(2p-1\right)/n}$ for $p=1,2,\dots ,n$(fold) \cite{R20}. If $\mathcal{R}$ stands for rotation matrix for both electron and hole, a BdG Hamiltonian is rotation-invariant if it satisfies \cite{R28}
\begin{equation} \label{GrindEQ__14_} 
\overline{\mathcal{R}}\,\overline{\mathcal{V}}H_{BdG}(\boldsymbol{k}){\overline{\mathcal{V}}}^+\,{\overline{\mathcal{R}}}^+=H_{BdG}(\mathcal{R}\boldsymbol{k}) 
\end{equation} 
where, $\overline{\mathcal{R}}=$diag ($\mathcal{R},{\mathcal{R}}^*$), $\overline{\mathcal{V}}=$diag (1, $e^{-i\phi }$) , and  $\phi =2\pi /3$. The elements ($\mathcal{R},\ \mathrm{1}$) and (${\mathcal{R}}^*, e^{-i\phi }$) act on electron and hole, respectively \cite{R28}. Also, $\mathcal{R}=\mathrm{exp}\mathrm{}(i{\sigma }_z\pi /3)$ for ${\sigma }_z$ on spin\cite{R28}. However
\begin{equation} \label{GrindEQ__15_} 
\mathcal{R}\left( \begin{array}{c}
c^{e,+}_{k,\uparrow } \\ 
c^{e,+}_{k,\downarrow } \end{array}
\right)=\left( \begin{array}{cc}
e^{i\pi /3} & 0 \\ 
0 & e^{-i\pi /3} \end{array}
\right)\left( \begin{array}{c}
c^{e,+}_{k,\uparrow } \\ 
c^{e,+}_{k,\downarrow } \end{array}
\right)=\left( \begin{array}{c}
{e^{i\pi /3}c}^{e,+}_{k,\uparrow } \\ 
e^{-i\pi /3}c^{e,+}_{k,\downarrow } \end{array}
\right) 
\end{equation} 
That is $\left( \begin{array}{c}
c^{e,+}_{k,\uparrow } \\ 
c^{e,+}_{k,\downarrow } \end{array}
\right)\to \left( \begin{array}{c}
{e^{i\pi /3}c}^{e,+}_{k,\uparrow } \\ 
e^{-i\pi /3}c^{e,+}_{k,\downarrow } \end{array}
\right)$ for electrons, and therefore for holes $\left( \begin{array}{c}
c^e_{-k,\downarrow } \\ 
c^e_{-k,\uparrow } \end{array}
\right)\to \left( \begin{array}{c}
{e^{i(\frac{\pi }{3}-\phi )}c}^e_{-k,\downarrow } \\ 
e^{i(-\frac{\pi }{3}-\phi )}c^e_{-k,\uparrow } \end{array}
\right)=\left( \begin{array}{c}
{e^{-i\pi /3}c}^e_{-k,\downarrow } \\ 
e^{-i\pi }c^e_{-k,\uparrow } \end{array}
\right)$ \cite{R28}. By considering the diagonal Hamiltonian diag$(E_1,E_2,E_3,E_4)$ at $K$- point, if $\left|\mu \right|>\left|{\beta }_{so}\right|$ then the eigenvalues $E_1$ and $E_2$ ($E^\prime_1$ and $E^\prime_2$) are negative, and the eigenvalues $E_3$ and $E_4$($E^\prime_3$ and $E^\prime_4$) are positive. But, if $\left|\mu \right|<\left|{\beta }_{so}\right|$ then the eigenvalues $E_2$ and $E_3$ ($E^\prime_1$ and $E^\prime_4$) are negative, and the eigenvalues $E_1$and $E_4$ ($E^\prime_2$, and $E^\prime_3$) are positive. Therefore, for $\left|\mu \right|>\left|{\beta }_{so}\right|$ the rotation eigenvalues (${\eta }_i,\ i=1,2,3,4$) are $e^{-i\pi /3}$, $e^{-i\pi }$, $e^{i\pi /3}$ , and $e^{-i\pi /3}$ which are related to $E_1$ , $E_2$,$\ E_3$ and $E_4$, respectively at $K$- point. It means that $\overline{\mathcal{R}}\,\overline{\mathcal{V}}=$diag($e^{-i\pi /3}$, $e^{-i\pi }$, $e^{i\pi /3}$, $e^{-i\pi /3}$)\cite{R28}. Also, for$\left|\mu \right|<\left|{\beta }_{so}\right|$, they are  $e^{i\pi /3}$, $e^{-i\pi }$, $e^{-i\pi /3}$ , and $e^{-i\pi /3}$ at $K$- point, and in consequence, $\overline{\mathcal{R}}\,\overline{\mathcal{V}}=$diag($e^{i\pi /3}$, $e^{-i\pi }$, $e^{-i\pi /3}$, $e^{-i\pi /3}$)\cite{R28}. Similarly, it can be shown that at $K^\prime$-point, for $\left|\mu \right|>\left|{\beta }_{so}\right|$ , $\overline{\mathcal{R}}\,{\overline{\mathcal{V}}}^+=$diag($e^{-i\pi /3}$, $e^{-i\pi }$, $e^{i\pi /3}$, $e^{-i\pi /3}$) and for $\left|\mu \right|<\left|{\beta }_{so}\right|$, $\overline{\mathcal{R}}\,{\overline{\mathcal{V}}}^+=$diag($e^{i\pi /3}$, $e^{-i\pi }$, $e^{i\pi /3}$, $e^{-i\pi /3}$)\cite{R28} if the diagonal Hamiltonian is written as diag$(E^\prime_1,E^\prime_2,E^\prime_3,E^\prime_4)$. But, the Chern number ($C$) in the three-fold symmetric system can be written as \cite{R20,R28}
\begin{equation} \label{GrindEQ__16_} 
e^{-i2\pi C/3}=\prod_{i\in 0cc.}{{\eta }_i(K){\eta }_i(K^\prime)} 
\end{equation} 
As a result, for $\left|\mu \right|>\left|{\beta }_{so}\right|$
\begin{widetext}
\begin{equation} \label{GrindEQ__17_} 
E_1=E^\prime_2=-\mu +{\beta }_{so}<0\to  e^{-i2\pi C/3}=e^{-i\pi /3}.\ e^{-i\pi }=e^{-i4\pi /3}\to C=2 
\end{equation}
\end{widetext}
\begin{widetext} 
\begin{equation} \label{GrindEQ__18_} 
E_2=E^\prime_1=-\mu -{\beta }_{so}<0\to  e^{-i2\pi C/3}=e^{-i\pi }.\ e^{-i\pi /3}=e^{-i4\pi /3}\to C=2 
\end{equation}
\end{widetext} 
Similarly, for $\left|\mu \right|<\left|{\beta }_{so}\right|$
\begin{widetext}
\begin{equation} \label{GrindEQ__19_} 
E_2=E^\prime_1=-\mu -{\beta }_{so}<0\to  e^{-i2\pi C/3}=e^{-i\pi }.\ e^{+i\pi /3}=e^{-i2\pi /3}\to C=1 
\end{equation} 
\end{widetext}
\begin{widetext}
\begin{equation} \label{GrindEQ__20_} 
E_3=E^\prime_4=\mu -{\beta }_{so}<0\to  e^{-i2\pi C/3}=e^{-i\pi /3}.\ e^{-i\pi /3}=e^{-i2\pi /3}\to C=1 
\end{equation}
\end{widetext} 
However, for two-dimensional materials, it has been shown that \cite{R30}
\begin{equation} \label{GrindEQ__21_} 
C_{2D}=\sum_{n\epsilon occ.}{C^{(n)}} 
\end{equation} 
Therefore, for $\left|\mu \right|>\left|{\beta }_{so}\right|$ , $C_{2D}=4$, for $\left|\mu \right|<\left|{\beta }_{so}\right|,$ $C_{2D}=2$, otherwise it is equal to zero\cite{R8}. \\

Let us consider a nanoribbon of Mo-atoms with zigzag edge as shown in Fig.1(a) and use Eq. (10) (in real-space) and Eq. (11), for finding the energy dispersion curve ($E(\boldsymbol{k})$).  Fig.3(a) shows the energy dispersion curve of the nanoribbon. As it shows, the valence and conduction bands intersect with each other and since the DOS at zero energy has significant value (Fig.3(b)), there are surface states and the ribbon behaves as a metal. It should be noted that, RSOC is not considered, here. Since we assume ${\beta }_{so}=60$ meV\cite{R27}, the difference between spin up and down electrons is negligible, and it is not shown.\\

Using the model of zigzag nanoribbon described in Eq. (12) and Eq. (13) the energy dispersion curve of the Bogoliubov-de Genes (BdG) Hamiltonian is found, without RSOC. Fig. 4(a) and Fig. 4(b) show the $E(\boldsymbol{k})$-curve for ${\beta }_{so}>{\Delta }_o$ and ${\beta }_{so}<{\Delta }_o$. As Fig. 4(a) shows, there are four zero energy states when ${\beta }_{so}>{\Delta }_o$ while it decreases to two for  ${\beta }_{so}<{\Delta }_o$. In addition, Fig. 4(c) shows that there are zero energy states when ${\beta }_{so}={\Delta }_o$. Therefore, it can be concluded that the model can predict the same topological properties which were found by using the Eq. (16) to Eq. (21), above. The zero energy states are referred to the Majorana states. 
\begin{figure}
\captionsetup{singlelinecheck = false, justification=raggedright}
{
    \includegraphics[width=\columnwidth]{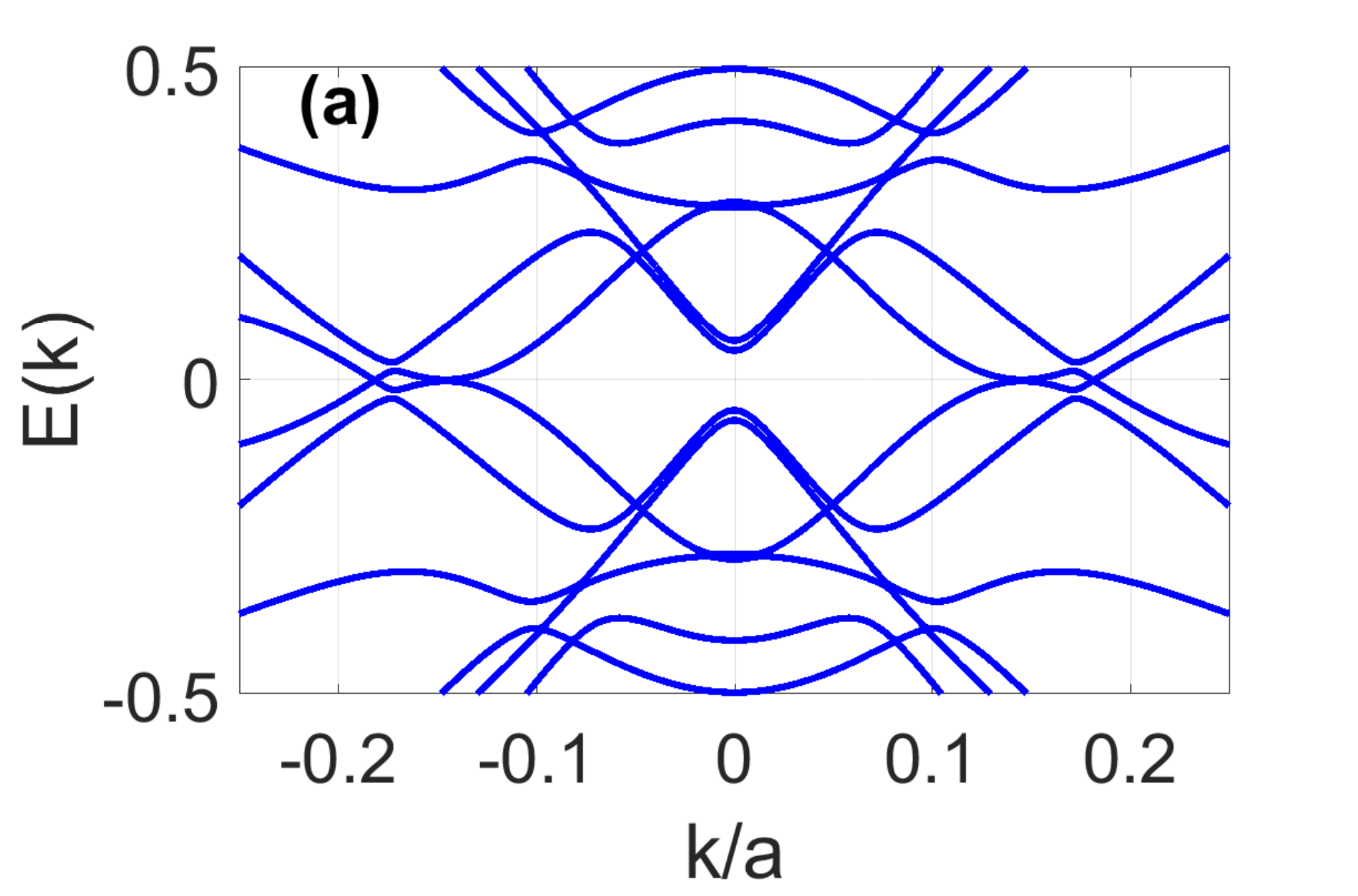}
}
\\

{
    \includegraphics[width=\columnwidth]{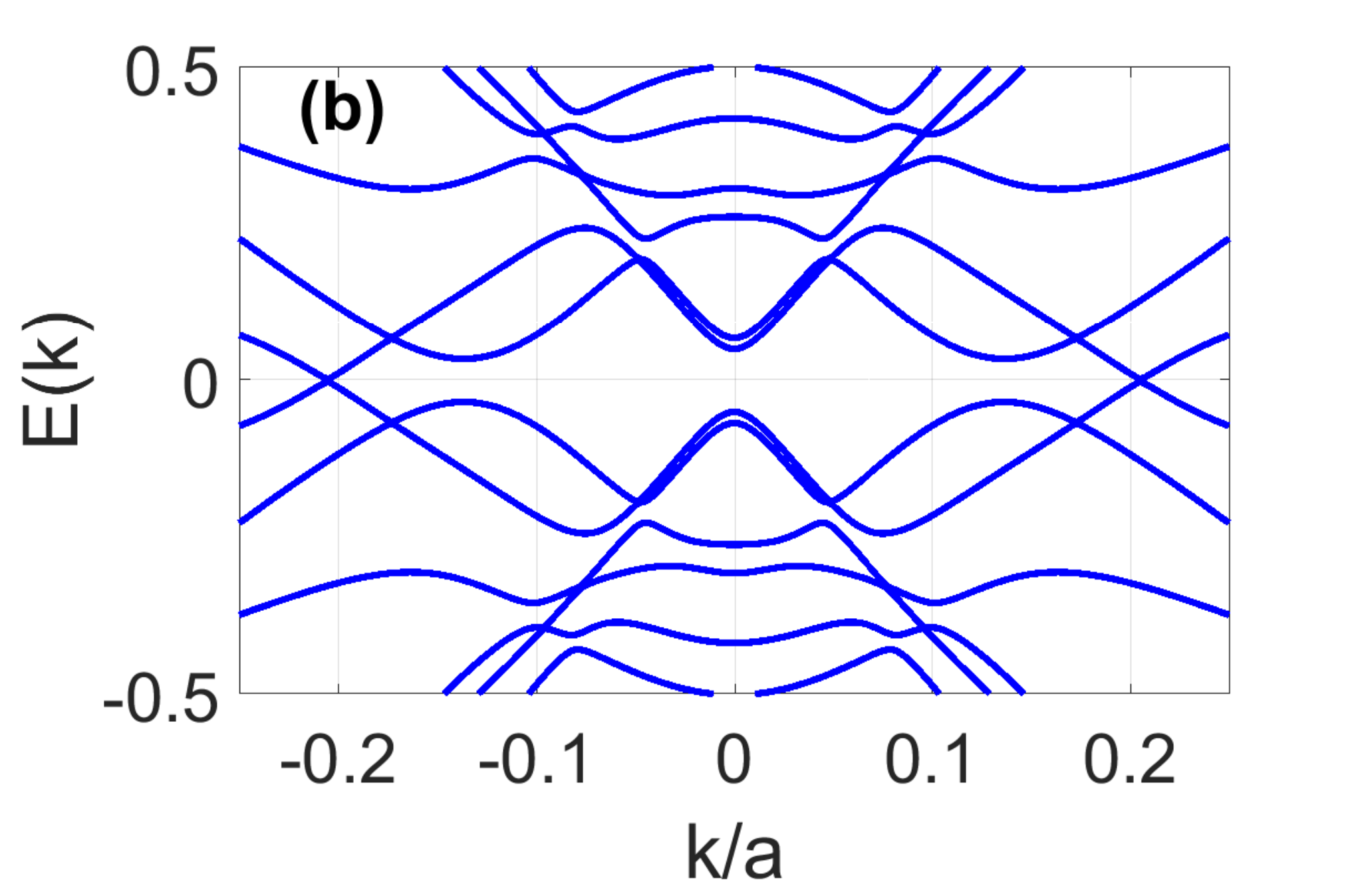}
}
\\

{
    \includegraphics[width=\columnwidth]{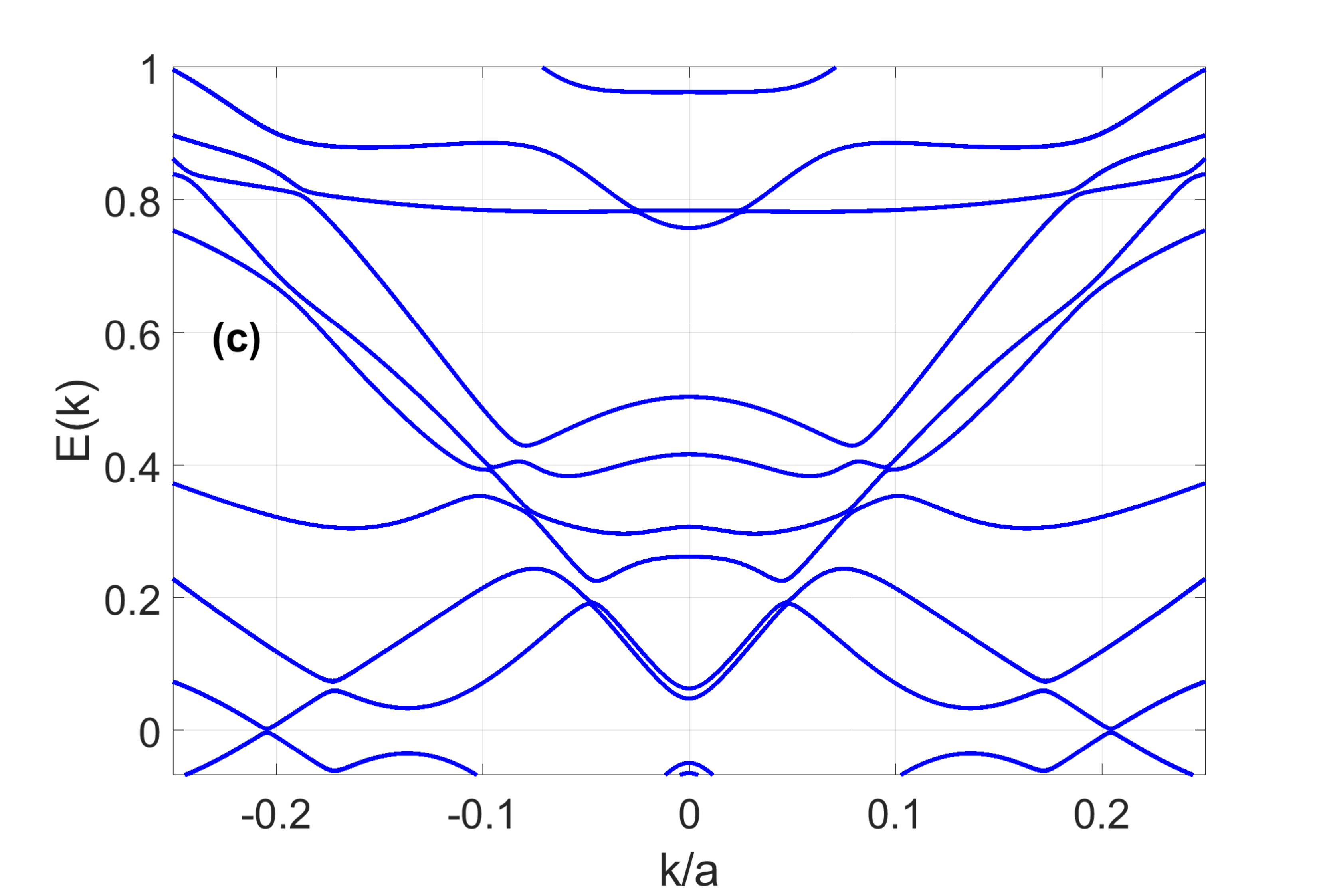}
}
\caption{Energy dispersion curve of Bogoliubov-de Genes (BdG) Hamiltonian (a) ${\beta}_{so}=60$ meV and ${\Delta}_{0}=6$ meV, (b) ${\beta}_{so}=6$ meV and ${\Delta}_{0}=60$ meV, and (c) ${\beta}_{so}=60$ meV and ${\Delta}_{0}=60$ meV. The $RSOC=0$, here. Each supercell includes nine Mo atoms.}
\end{figure}
\\

Here, the simpler model which was introduced at the end paragraph of section III is used to study the effect of chemical energy in respect to the ISOC strength. Fig. 5(a) shows the energy dispersion curve when ${\beta }_{so}=60$ meV,  ${\Delta }_0=6$ meV, and $\mu =87$ meV \cite{R28}. As it shows, since $\mu >{\beta }_{so}$, there are four zero energy states while for $\mu =57$ meV, there are two zero energy states (Fig. 5(b)). The model shows that when $\mu \gg {\beta }_{so}$, a gap is opened in the curve (it is not shown in figures). Finally, as Fig. 6 shows, the DOS has significant values at zero energy for both cases i.e., $\mu >{\beta }_{so}$ and $\mu <{\beta }_{so}$. Therefore, the simple model can explain the analytical results qualitatively (at least).
\begin{figure}
\captionsetup{singlelinecheck = false, justification=raggedright}
{
    \includegraphics[width=\columnwidth]{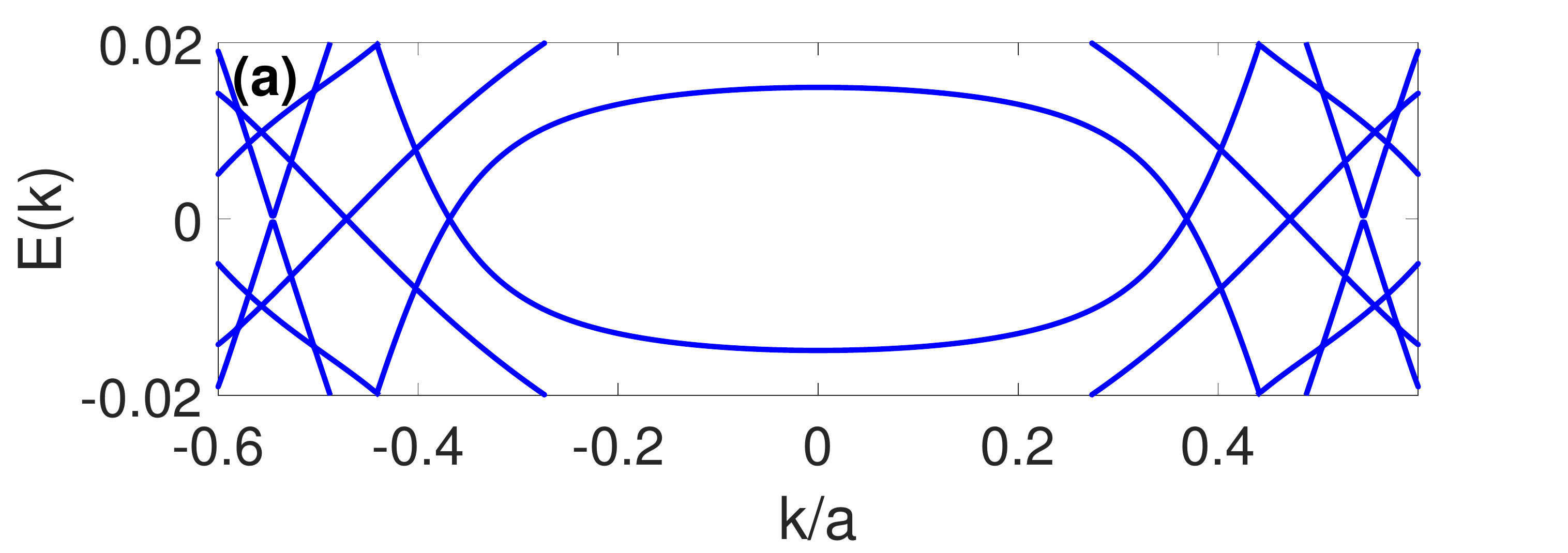}
}
\\

{
    \includegraphics[width=\columnwidth]{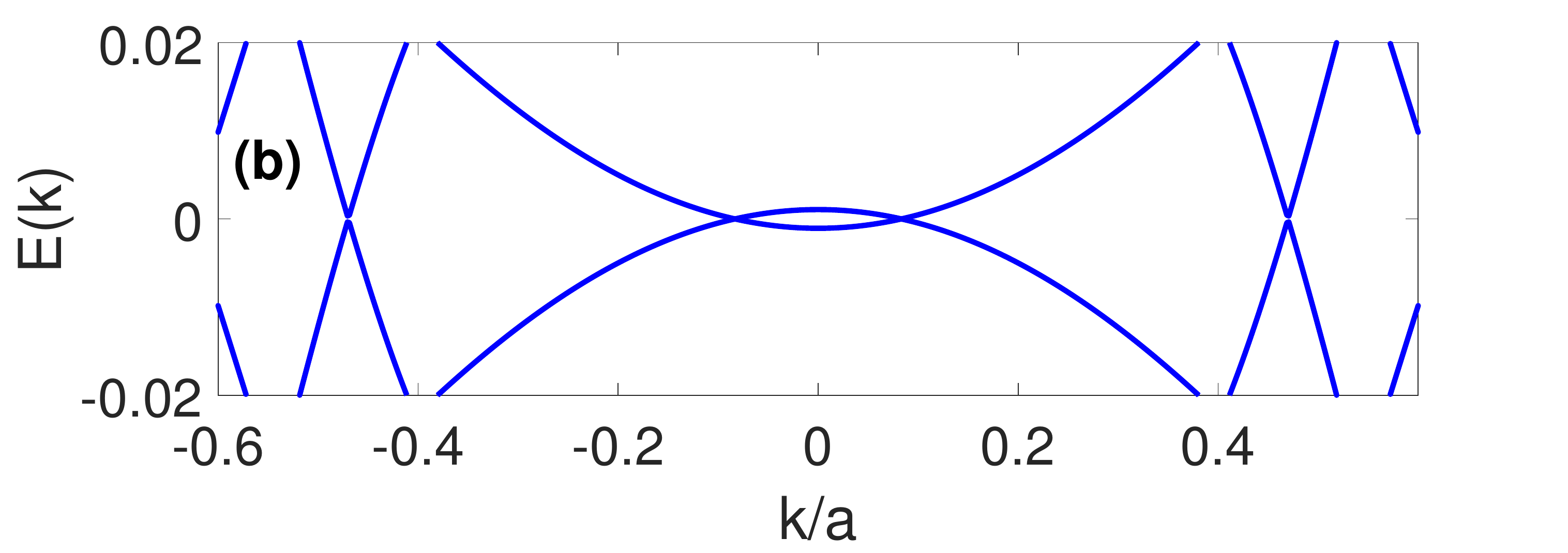}
}

\caption{Energy dispersion curve of Bogoliubov-de Genes (BdG) Hamiltonian (a) $\mu=87$ meV and (b) $\mu=57$ meV. Here,  ${\beta}_{so}=60$ meV , ${\Delta}_{0}=6$ meV and $RSOC$ strength is equal to zero. Each supercell includes nine Mo atoms.}
\end{figure}
\begin{figure}
\captionsetup{singlelinecheck = false, justification=raggedright}
{
    \includegraphics[width=\columnwidth]{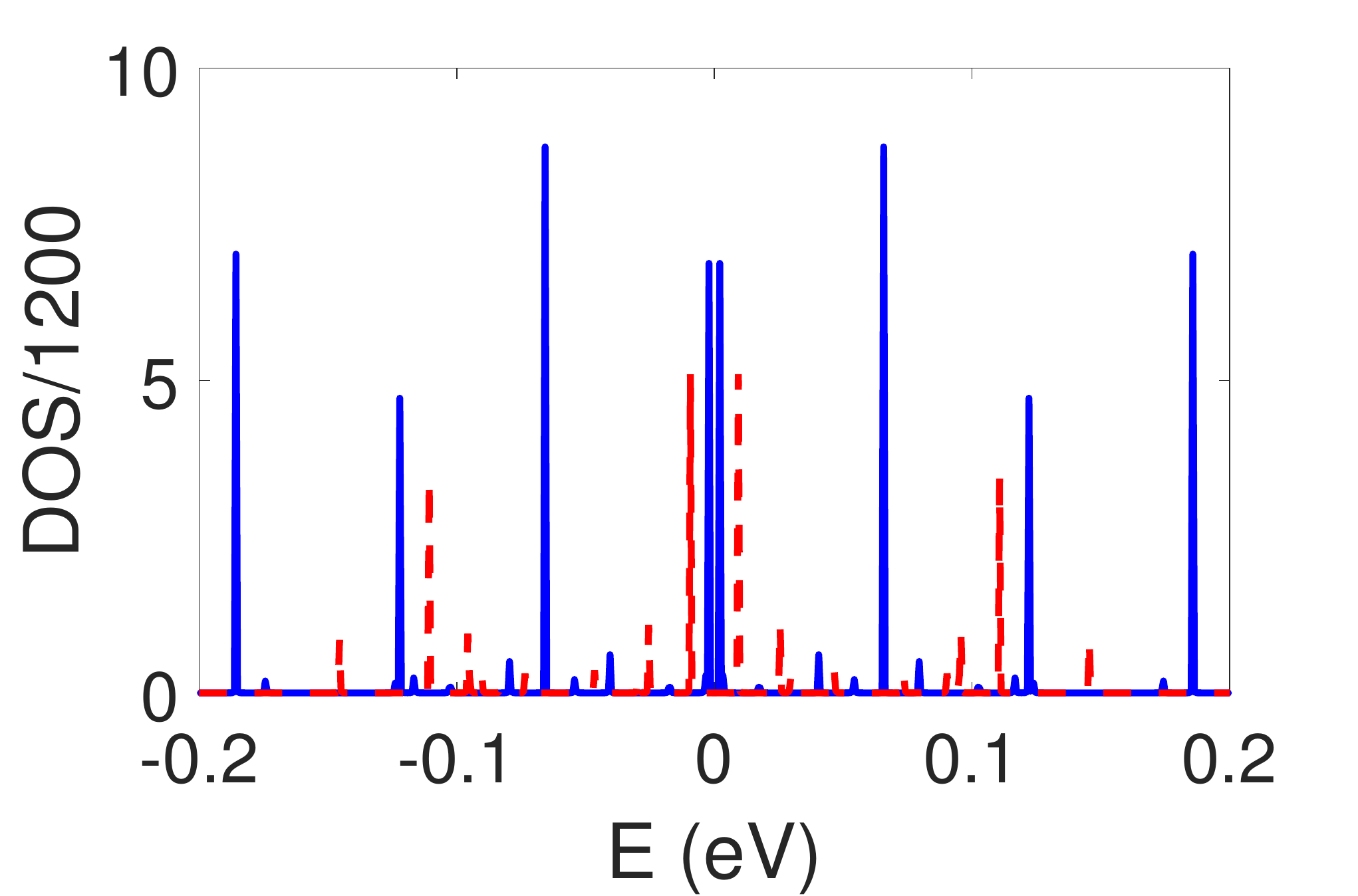}
}

\caption{(Color online) Density of states ($DOS$) versus energy for $\mu>{\beta}_{so}$ in filled blue color and for $\mu<{\beta}_{so}$ in dashed red color. Here, $RSOC$ strength is equal to zero. Each supercell includes nine Mo atoms. }
\end{figure}
\\

Finally, the effect of mechanical transformation on the topological properties of monolayer MoS${}_{2}$ is investigated. An uniaxial tension field ($\boldsymbol{T}$) is applied to the Mo-plane. The vector $\boldsymbol{T}$ has the angle $\theta $ with the $x$-axis which is in parallel with the zigzag edge of the nanoribbon. It can be shown that the strain tensor ($\boldsymbol{\varepsilon }$) can be written as \cite{R31}
\begin{equation} \label{GrindEQ__22_} 
\boldsymbol{\varepsilon }=\tau \left( \begin{array}{ccc}
{cos}^2\theta -v_{\parallel }{sin}^2\theta  & \left(1+v_{\parallel }\right)sin\theta cos\theta  & 0 \\ 
\left(1+v_{\parallel }\right)sin\theta cos\theta  & {sin}^2\theta -v_{\parallel }{cos}^2\theta  & 0 \\ 
0 & 0 & -v_{\bot } \end{array}
\right) 
\end{equation} 
where, $\tau $ tension strain, $v_{\parallel }$ and $v_{\bot }$ are in-plane and out-of-plane Poisson's ratios, respectively. Therefore,
\begin{equation} \label{GrindEQ__23_} 
D=Tr\left[{\epsilon }_{ij}\right]=\tau (1-v_{\parallel }-v_{\bot }) 
\end{equation} 
and for $\theta =0$ , i.e., $\boldsymbol{T}$ is parallel to the zigzag edge                                         
\begin{equation} \label{GrindEQ__25_} 
F^{\epsilon }=({\epsilon }_{yy}-{\epsilon }_{xx}+i\epsilon 2{\epsilon }_{xy})=\tau (-1+v_{\parallel }) 
\end{equation} 
and for $\theta =\pi /2$, i.e., $\boldsymbol{T}$ is perpendicular to the zigzag edge
\begin{equation}
F^{\epsilon }=({\epsilon }_{yy}-{\epsilon }_{xx}+i\epsilon 2{\epsilon }_{xy})=\tau (1-v_{\parallel }) 
\end{equation}
By using the values of all constants at Eq. (9) which provided before \cite{R13,R14,R15}, it can be shown that 
\begin{equation} \label{GrindEQ__25_} 
\beta {\left|{\eta }_2F^{\epsilon }\right|}^2+{\delta }_1D+{\delta }_3D^2+s_zD\left(\delta {\beta }_{so1}\right)=-0.0844 +0.07s_z
\end{equation} 
\begin{equation} \label{GrindEQ__26_} 
\alpha {\left|{\eta }_3F^{\epsilon }\right|}^2++{\delta }_2D+{\delta }_4D^2+s_zD\left(\delta {\beta }_{so2}\right)=-0.4295-0.0142s_z 
\end{equation} 
and

\begin{equation}
{\eta }_1F^{\epsilon }+\kappa {\eta }^4_4{F^{\epsilon }}^2=-7.58\times {10}^{-4}  
\end{equation}
which is negligible.
\noindent These values should be added to Eqs. (17-20). However, as we assumed that the tension field does not change the rotation symmetry (i.e., it is small), these equations are satisfied, and in consequence, the topological properties are preserved under the small uniaxial tensions field. 

It should be noted that one can calculate the effect of biaxial strain, similarly. For example, under trigonal deformation such as $(u_{x},u_{y} )={\frac{u_{0}}{2}}(xy,{\frac{x^{2}-y^{2}}{2}})$, by using Eq. (8), one can easily show that $F^{\epsilon }=u_{0} (y,-x)$. Also, for arc-shape deformation $(u_{x},u_{y} )=({\frac{xy}{R}},{\frac{-x^{2}}{2R}})$ where R is arc radius, $F^{\epsilon }=({\frac{y}{R}},0)$. But, the deformed new Hamiltonian is now given by the Eq. (50) of Ref. \cite{R15} and the effect of curvature and gauge fields should be considered. Finally, the effect of deformation on the topological properties can be studied by calculating the amount of each element of the new Hamiltonian, if the deformation does not change the symmetry of the lattice. \\
As the electronic properties of two dimensional materials are studied near $K$ and $K^\prime$ points in this research, the behavior of energy dispersion curve near these points is considered. But,for studying the effect of other orbitals, complete Hamiltonian i.e., Eq. (2) of Ref. \cite{R29} should be considered. Then, Eq. (34) to Eq. (37) of Ref. \cite{R15} can be used to calculate the effect of deformation. Obviously, the calculation method  is more complicated than the provided method in this article and needs more justification.\\
For whom might be interested in the effect of doping can be only studied by using the ab-inito methods such as density functional theory, which is out of the scope of the article.
\section{Summary}\label{sec:summary}
It has been shown that the $4d_{z^2}$ orbital is the dominant component of the states near the CBM and the VBM located at $K$ and $K^\prime$ points in the monolayer MoS${}_{2}$. We have considered the triangular lattice of Mo-plane and studied the topological superconductivity in monolayer MoS${}_{2}$. By considering the spin-singlet pairing at $K$ and $K^\prime$ points, we have shown that the BdG-Hamiltonian matrix is diagonal and has four distinct eigenvalues which are functions of chemical energy and ISOC strength. Using the rotation symmetry of the fixed point $K$ and $K^\prime,$ it has been shown that the Chern number is equal to four (two) when the chemical potential ($\mu $) is greater (smaller) than ISOC strength (${\beta }_{so}$) and otherwise it is equal to zero. Also, we have introduced two simple tight-binding BdG-Hamiltonian models for finding the zero energy states (i.e., Majorana states) and confirming the analytical results. In the first model, the ISOC was considered by choosing the $d_{x^2-y^2}$ and $d_{xy}$ orbitals of Mo-atoms. We have shown that when ${\beta }_{so}$ is greater than pairing potential (${\Delta }_0$) there are four zero energy states. Under the same condition and using the second single-band tight-binding Hamiltonian, it has been shown that for both  $\mu >{\beta }_{so}$ and $\mu <{\beta }_{so}$ there are zero energy states and for $\mu \gg {\beta }_{so}$, a gap is opened in the energy dispersion curve. Finally, we have shown that under small uniaxial strain which can be parallel or perpendicular to the zigzag edge of Mo-nanoribbon, the topological properties are preserved.

\nocite{*}
% Create the reference section using BibTeX:
\bibliography{article}

\end{document}